# Spin Hall magnetoresistance as a probe for surface magnetization in Pt/CoFe$_2$O$_4$ bilayers

Miren Isasa[1], Saül Vélez[1], Edurne Sagasta[1], Amilcar Bedoya-Pinto[1], Nico Dix[2], Florencio Sánchez[2], Luis E. Hueso[1,3], Josep Fontcuberta[2] and Fèlix Casanova[1,3]

[1]CIC nanoGUNE, 20018 Donostia-San Sebastian, Basque Country, Spain.
[2]Institut de Ciència de Materials de Barcelona (ICMAB-CSIC), Campus UAB, 08193 Bellaterra, Catalonia, Spain.
[3]IKERBASQUE, Basque Foundation for Science, 48013 Bilbao, Basque Country, Spain.

**Abstract**

We study the spin Hall magnetoresistance (SMR) in Pt grown *in situ* on CoFe$_2$O$_4$ (CFO) ferrimagnetic insulating (FMI) films. A careful analysis of the angle-dependent and field-dependent longitudinal magnetoresistance indicates that the SMR contains a contribution that does not follow the bulk magnetization of CFO but it is a fingerprint of the complex magnetism at the surface of the CFO layer, thus signaling SMR as a tool for mapping surface magnetization. A systematic study of the SMR for different temperatures and CFO thicknesses gives us information impossible to obtain with any standard magnetometry technique. On one hand, surface magnetization behaves independently of the CFO thickness and does not saturate up to high fields, evidencing that the surface has its own anisotropy. On the other hand, characteristic zero-field magnetization steps are not present at the surface while they are relevant in the bulk, strongly suggesting that antiphase boundaries are the responsible of such intriguing features. In addition, a contribution from ordinary magnetoresistance of Pt is identified, which is only distinguishable due to the low resistivity of the *in-situ* grown Pt.

## I. Introduction

Spin-orbitronics is a novel direction of spintronics which exploits the strong spin-orbit coupling (SOC) present in non-magnetic (NM) metals and semiconductors for the generation, manipulation and detection of pure spin currents [1]. Many different phenomena arising from spin-orbit coupling, such as magnetic skyrmions [2-4], the Rashba and Dresselhaus effects [5-7], the spin Hall effect (SHE) [8,9] or the spin-orbit torques for magnetic switching of ferromagnetic elements [10,11] are intensively being explored. Of particular interest is the SHE, a phenomenon in which an unpolarized charge current flowing through a NM metal with strong SOC is converted into a transverse spin current due to the opposite scattering of spin-up and spin-down electrons [8,9]. According to Onsager's reciprocity relations, a spin current flowing through a NM metal with strong SOC will in turn create a transverse charge current, known as the inverse SHE (ISHE). Extensive work has been carried out to quantify the strength of the effect in different metals [12-20].

Recently, a new type of magnetoresistance, which combines the SHE and ISHE, has been discovered: the spin Hall magnetoresistance (SMR) [21-27]. This new effect appears in bilayers formed by a NM metal with a strong SOC and a ferromagnetic insulator (FMI). Nakayama and coworkers [22] observed that depending on the magnetization direction of the FMI, the spin current created by the SHE on the NM could be absorbed, *via* spin transfer torque, or reflected at the NM/FMI interface, creating a change in resistance in the NM. Given the interfacial origin of the effect, controlling the quality of the NM/FMI interface is crucial to get an enhanced magnetoresistance effect [27, 28]. Although only recently discovered, the SMR has already proven to be a successful approach to quantify the spin-mixing interfacial conductance of NM/FMI bilayers [24,25,27,28], a concept at the base of this and other spin-dependent phenomena such as the spin Seebeck effect [29-32], the spin pumping [29,33-35] or the magnetic gating of pure spin currents [36,37].

In this work, we demonstrate a novel functionality of the SMR by sensing the surface magnetization of the FMI. For this purpose, we have chosen $CoFe_2O_4$ (CFO), a room-temperature ferrimagnetic insulating oxide whose surface magnetization differs from that of the bulk due to its complex atomic constitution, competing magnetic interactions and symmetry breaking [38,39]. The presence of $Co^{2+}$ ions in its cubic spinel structure, $(Fe^{3+})[Co^{2+}Fe^{3+}]O_4$, anticipates a large magnetic anisotropy in CFO [38] and the competing nature of magnetic interactions in spinels may lead to different magnetic properties at the surface [39]. As NM metal we choose Pt, the metal most commonly used in experiments involving the spin-mixing conductance of NM/FMI bilayers [22-35]. In order to obtain clean Pt/CFO interfaces, Pt and CFO layers were grown *in situ*, in a single process, without air exposure between Pt and CFO layer deposition. We report magnetoresistance measurements, displaying features fully compatible with SMR, but with an additional signal arising from the ordinary magnetoresistance (OMR) of Pt. This is made distinguishable by in-situ growth of bilayer. More importantly, by studying the field dependence of the longitudinal magnetoresistance arising from SMR, we are able to resolve the distinct surface magnetization behavior of the CFO films, compared to its bulk magnetization.

**II. Experimental details**

A first set of three 40-nm-thick CFO films were epitaxially grown on (001) $SrTiO_3$ (STO) substrates by pulsed laser deposition. The beam of a KrF excimer laser was focused on a CFO target at an angle of 45º. The fluence and repetition rate, were 1.5(3) $J/cm^2$ and 5 Hz, respectively. The substrate temperature was about 550 °C and oxygen pressure $P_{O2}=$ 0.1 mbar [40]. Crucial for SMR phenomenon is the nature of the interface between the FMI and the NM (Pt in the present case) layer [23,27,28,41-43]. As shown in a previous work [27] the magnitude of the SMR strongly depends on the interface preparation conditions, being optimal when the NM is grown *in situ* on top of the FMI. For this reason, Pt with three different thicknesses ($t_{Pt}=$ 6.5, 4 and 2 nm) was epitaxially grown by dc sputtering on CFO by an *in-situ* process at 400 ºC. The thicknesses of the CFO and Pt layers were inferred from growth rate calibration by X-ray reflectometry. θ/2θ scans were used to confirm that all CFO films were fully (001) textured without spurious phase and that Pt layers were also (001) textured. For transport measurements the Pt layers were patterned into Hall bars (width W=100 μm and length L=800 μm), as sketched in Fig. 1, using electron-beam lithography with negative resist on top of the Pt layer, followed by Ar-ion milling and lift-off. The longitudinal base resistances ($R_{L0}$) of the 6.5-, 4- and 2-nm-thick Pt films are 272 (144) Ω, 403 (364) Ω and 1440 (1198) Ω, respectively, yielding resistivities of 21.5 (11.4), 20.2 (17.7) and 36.0 (30.0) μΩcm at 300 (50) K.

A second set of four samples was fabricated in the same way. For this set, 20-, 30-, 40- and 60-nm-thick CFO films were grown, whereas Pt thickness was fixed to 2 nm.

Magnetization and magnetotransport measurements were performed in the same liquid-He cryostat, where the temperature varied from 300 K to 40 K with externally applied magnetic fields (**H**) ranging from -9 T to 9 T. Vibrating sample magnetometry (VSM) was used to determine the magnetization of the CFO films. Transport measurements were performed using a Keithley 6221 sourcemeter and a Keithley 2182A nanovoltmeter operating in the dc-reversal method [44] with 100 μA of applied current. Field dependent magnetoresistance (FDMR) measurements were carried out in the longitudinal configuration, as sketched in Fig. 1, where the field was applied in three different directions: (i) along the current direction (**j** direction), (ii) in-plane and transverse to **j** (**t** direction) and (iii) out-of-plane (**n** direction). Angle dependent magnetoresistance (ADMR) measurements were also performed in the longitudinal configuration, in which the applied field was fixed at 9 T and varied along the different **H**-rotation planes. Their corresponding angles are defined as: (i) α (from **j**, angle α=0, towards **t**), (ii) β (from **n**, angle β=0, towards **t**) and (iii) γ (from **n**, angle γ=0, towards **j**). All these **H**-rotation planes are schematically shown in Fig. 1.

**III. Results and discussion**

**IIIa. Angle dependent magnetoresistance measurements**

According to the SMR theory, the angular dependence of the measured longitudinal ($\rho_L$) and transverse ($\rho_T$) resistivity is given by [21]:

$$\rho_L = \rho_0 + \Delta\rho_1(1 - m_t^2) \quad (1)$$
$$\rho_T = \Delta\rho_1 m_j m_t + \Delta\rho_2 m_n \quad (2)$$

where $\rho_0$ is the baseline resistivity of the NM layer, the ratio $\Delta\rho_1/\rho_0$ is the SMR and $\Delta\rho_2$ accounts for an anomalous Hall-like contribution. **m**($m_j$, $m_t$, $m_n$)=**M**/$M_s$ are the cosine directors of the magnetization **M** along the **j**, **t** and **n** directions. $M_s$ is the saturation magnetization of CFO. In Fig. 1, we show the measured ADMR of the Pt(6.5 nm)/CFO(40 nm) sample at 300 K, defined by their corresponding angles α, β and γ. All the measurements have been performed in the longitudinal configuration at fields $\mu_0 H$ (9 T) much larger than the coercive field $\mu_0 H_C$ of the CFO (see Fig. 5a). Hence, we initially assume that, in Fig. 1, **m** roughly follows **H**. Transverse ADMR measurements as a function of angle α yield the same amplitude ($\Delta\rho_1$) as the longitudinal measurements [27], whereas the term $\Delta\rho_2$ is completely hidden by the ordinary Hall effect of Pt when **H** is rotated in the β and γ plane [25]. Since these measurements do not give additional information, they are not shown for the sake of simplicity.

Based on the SMR scenario, the longitudinal resistance $R_L$ of Pt should only change when the direction of the magnetization changes with respect to the spin polarization, **s**, which points to **t** direction due to the symmetry of the SHE [22]. Thus, a change in resistance should only appear when the field is rotated along α and β angles. The measured resistance follows $R_L(\alpha) \propto \cos^2(\alpha)$ and $R_L(\beta) \propto \cos^2(\beta)$, respectively, and the magnetoresistance value should be similar in both cases, $\Delta R(\beta) = \Delta R(\alpha)$. Additionally, $R_L(\gamma)$ should not vary when **H** is rotated along γ, as in this case **M** is always perpendicular to **s**, and so $m_t = 0$. However, the measurements in Fig. 1 reveal a different scenario. On the one hand, $R_L(\gamma)$ is not a constant value when varying H (Fig. 1a). On the other hand, the change in $R_L(\beta)$ (Fig. 1b) is different to the change in $R_L(\alpha)$ (Fig. 1c). In fact, the difference between both curves [$R_L(\alpha)$ and $R_L(\beta)$] yields the same modulation observed in $R_L(\gamma)$. A very controversial issue when placing the Pt next to a FMI is the magnetization that can be induced in Pt by proximity effect, since Pt is close to the Stoner ferromagnetic instability [45-48]. If this was the case, $R_L(\gamma)$ measurements, which follows $R_L(\gamma) \propto \cos^2(\gamma)$, could be a signature of the anisotropic magnetoresistance (AMR) of the magnetized Pt. As AMR is sensitive to the variation of the magnetization with respect to the charge current direction, it would also be contributing to the $R_L(\alpha)$ measurements, being this configuration sensitive to both AMR and SMR.

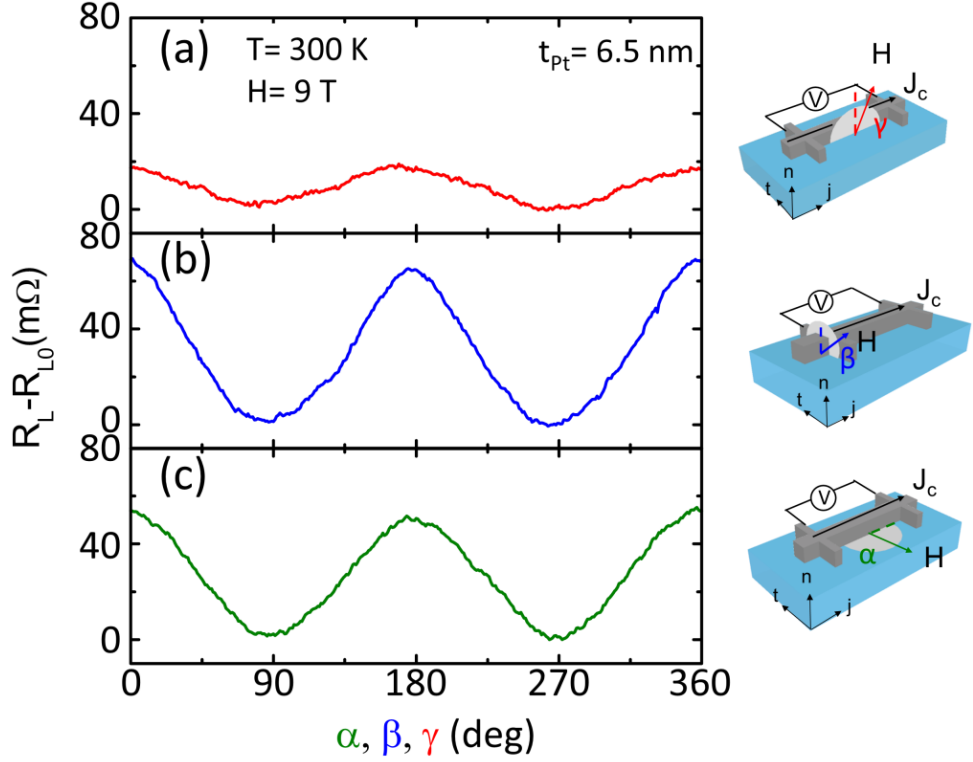

**Figure 1**. *Angle-dependent magnetoresistance measurements at 9 T and 300 K for the Pt(6.5 nm)/CFO(40 nm) sample along (a) γ, (b) β and (c) α rotation planes. $R_L$ is the measured longitudinal resistance and $R_{L0}$ the subtracted background. The sketches in the right define the Hall bar geometry, the longitudinal measurement set-up and the angles α, β and γ.*

To further understand the modulation that we observe in $R_L(\gamma)$, we performed the same measurements for samples with different Pt thicknesses and at different temperatures, ranging from 40 to 300 K (Fig. 2). As observed, the normalized magnetoresistance in $R_L(\gamma)$, $\Delta R(\gamma)/R_{L0}$, is present for different Pt thicknesses and for all temperatures, being largest for the case of the thickest Pt. The fact that this contribution decreases when reducing thickness rules out that it arises from a proximity effect which should be more relevant for thinner films [49]. Additionally, Fig. 2 shows that $\Delta R(\beta)/R_{L0}$, the purely SMR signal, becomes more important at low thicknesses, as expected [23,25,27].

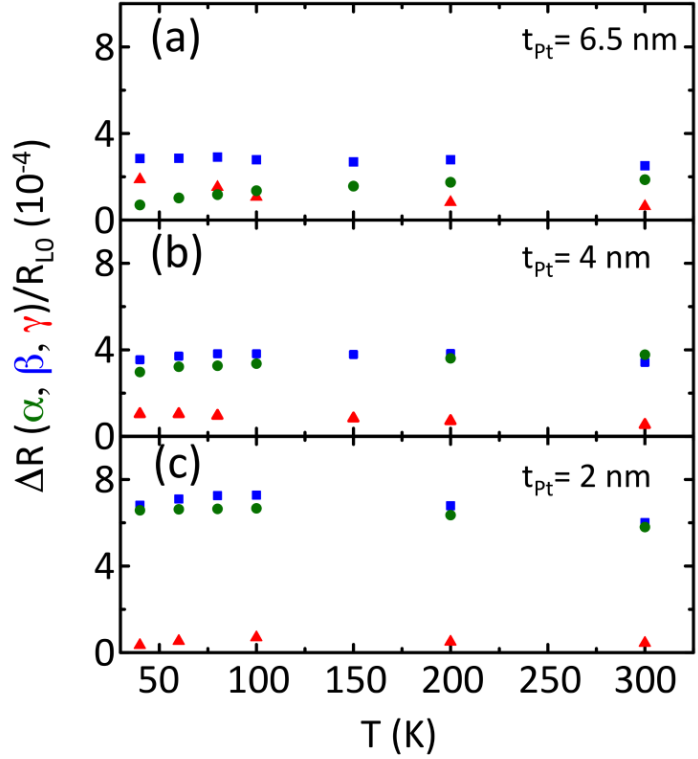

**Figure 2**. *The amplitude of ADMR at 9 T as a function of temperature, for the three different angles α (green circles), β (blue squares) and γ (red triangles), for (a) Pt(6.5 nm)/CFO(40 nm) (b) Pt(4 nm)/CFO(40 nm) and (c) Pt(2 nm)/CFO(40 nm) samples.*

### IIIb. Field dependent magnetoresistance measurements

Another strategy to rule out the possibility of having AMR in a magnetized Pt is performing FDMR measurements fixing the direction of the magnetic field (**n**, **t** or **j**) and sweeping it from -9 T to 9 T (see Fig. 3). If AMR was present in our samples, we should obtain the distinct trend of the magnetoresistance when the field is applied perpendicular to the charge current (**H**||**t** and **H**||**n**) and when it is applied parallel to the charge current (**H**||**j**). However, this is not what Fig. 3 shows, since the **H**||**n** curves have the same magnetoresistance trend as the **H**||**j** curves and opposite to the one observed with **H**||**t**, irrespective of the Pt thickness. Therefore, the AMR contribution should be discarded. This conclusion is in agreement with recent atomic selective magnetic measurements in similar Pt/CFO layers where, within the experimental resolution, no magnetic moment has been found at the Pt [50].

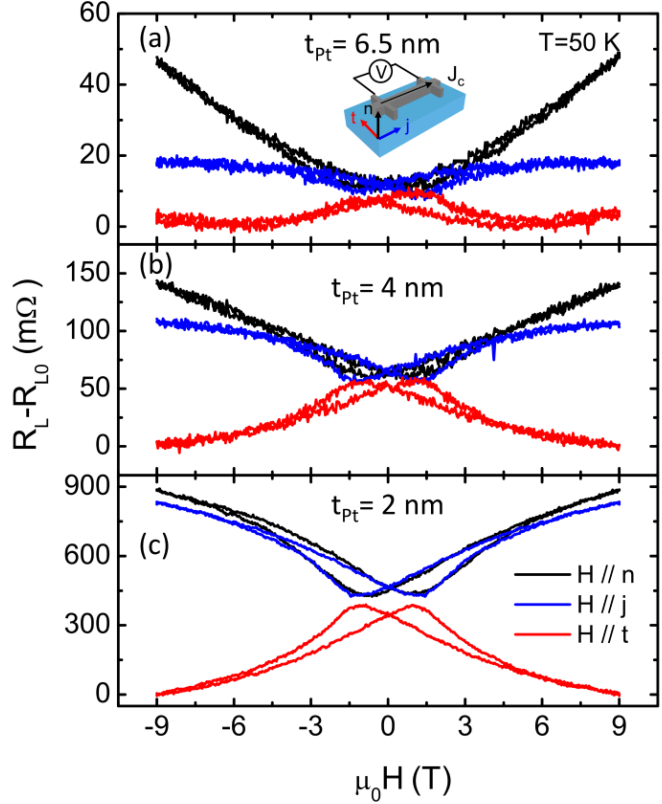

**Figure 3.** *Longitudinal resistance $R_L$ for (a) Pt(6.5 nm)/CFO(40 nm) (b) Pt(4 nm)/CFO(40 nm) and (c) Pt(2 nm)/CFO(40 nm) samples, as a function of H applied along **t** (red curves), **j** (blue curves) and **n** (black curves). $R_{L0}$ is the subtracted background. The orientations of the applied field and the measurement configuration are sketched in the inset. All measurements were done at 50 K.*

It is worth noting the different resistance values measured for **H**∥**j** and **H**∥**n** at high fields, which accounts for the modulation in $R_L(\gamma)$ and it strongly depends on the Pt thickness, being more pronounced for the thickest sample. A possible explanation to this behavior is related to the ordinary magnetoresistance (OMR) in Pt. This magnetoresistance effect appears in metals and semiconductors and it occurs because conduction electrons are displaced from their trajectories by the Lorentz force exerted by an externally applied magnetic field. The magnetoresistance due to the OMR ($\Delta R_{OMR}/R_{L0}$) can be described by the Kohler´s rule, which depends on the applied field and resistivity in the form of [51,52]:

$$\frac{\Delta R_{OMR}}{R_{L0}} = a \left(\frac{H}{\rho}\right)^n \tag{3}$$

where *a* and *n* are material dependent constants, with *n* between 1 and 2. This magnetoresistance should be characteristic of the Pt, but independent of the Pt thickness, as far as its scattering length is not affected by size effects. To verify this, we subtract $R_L(\mathbf{H}\|\mathbf{n}) - R_L(\mathbf{H}\|\mathbf{j})$ to obtain $\Delta R$ and normalize it to $R_{L0}$, which should correspond to the extra magnetoresistance present in the system. Subsequently, we plot the extra magnetoresistance as a function of H/ρ (see Fig. 4). As expected, all curves define a parabola, with $n=1.8$, confirming that this extra effect is OMR. OMR has not been detected in previous studies on Pt/FMI [25,53], due the large resistivity usually obtained in Pt thin films (~41-60 μΩ cm). Our Pt, grown *in situ* at 400ºC on top of epitaxial (001) CFO, is fully textured in the (001) direction, leading to lower resistivity and therefore to a non-negligible OMR contribution.

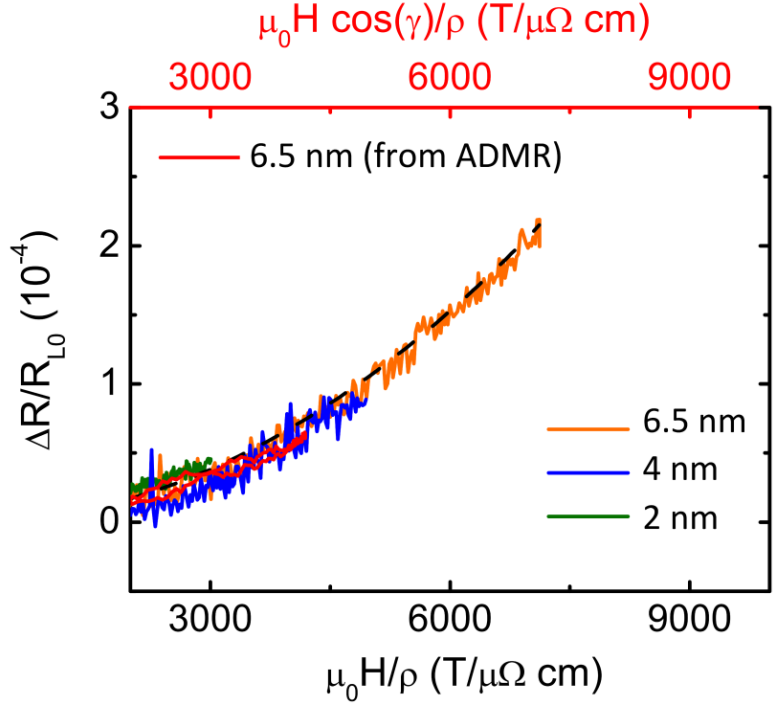

**Figure 4.** *Orange, blue and green curves are the additional magnetoresistance observed for out-of-plane H, $[R_L(\mathbf{H}||\mathbf{n})-R_L(\mathbf{H}||\mathbf{j})]/R_{L0}$, as a function of H/ρ for the three Pt/CFO(40 nm) samples with different Pt thicknesses at 50 K. The red curve is the longitudinal magnetoresistance $(R_L(\gamma) - R_{L0})/ R_{L0}$ as a function of the out-of-plane H for the 6.5-nm-thick Pt, at 300 K. The dashed black line is a guide for the eye.*

Note that, once the OMR contribution is identified in $R_L(\mathbf{H}||\mathbf{n})$, the curves arising solely from the SMR contribution are $R_L(\mathbf{H}||\mathbf{t})$ and $R_L(\mathbf{H}||\mathbf{j})$. As expected (Fig. 3), the curves show a mirror symmetry, have the same shape with the Pt thickness and the SMR magnitude decreases with increasing thickness [22,23,25].

Now we can safely ascribe the behavior observed in Fig. 1a to OMR, where the angular dependence comes from the out-of-plane component of the field (Hcos(γ)). If we plot the measured $R_L(\gamma)$ as a function of Hcos(γ) we can see that it nicely follows Kohler´s curve (red curve in Fig. 4). Note that the curves in Fig. 4 correspond to different temperatures, where FDMR measurements are done at 50 K and ADMR at 300 K. The fact that they all lie over the same curve is a clear evidence that OMR is temperature independent.

**IIIc. Comparison between VSM and SMR measurements**

Once we have identified the coexistence of SMR and OMR in our system, we move now to the comparison between the magnetic properties of the CFO thin films and the spin Hall magnetoresistance of the Pt/CFO bilayers. Figure 5a shows the hysteresis loops M(H) of the Pt(2 nm)/CFO(40 nm) sample obtained by VSM when applying the magnetic field **H** along **t** and **n** directions at 50 K. As can be seen from the hysteresis loop when the field is applied in plane, M(**H**||**t**) curve, the large coercive fields $\mu_0 H_c(\mathbf{t}) \approx \pm 1.2$ T and the fact that hysteresis only disappears at ≈ 5 T are signatures of the strong magnetic anisotropy typical of CFO thin films [54,55]. The shape of the hysteresis loop when the field is applied out of plane, M(**H**||**n**), indicates a harder magnetization axis and, correspondingly, the coercive field $\mu_0 H_c(\mathbf{n}) \approx \pm 0.44$ T and the magnetic remanence are smaller. The saturation magnetization ($M_s$=230 emu/cm$^3$) is lower than the corresponding bulk value as commonly observed in spinel thin films [56-59] and attributed to the presence of antiphase boundaries (APB) [56,57] or to surface anisotropy effects [39]. Characteristic steps are observed in the hysteresis loops around zero field. These step-like features are commonly found in CFO thin films [54,60] and other ferrimagnetic oxides such as $Fe_3O_4$ [57] or $\varepsilon\text{-}Fe_2O_3$ [61], and have been attributed to the result of coupled antiferromagnetic domains due to the presence of APBs by Sofin *et al.* [62]. A larger density of APBs would lead to a larger step at zero field. The diamagnetic background, arising mainly from the STO substrate, has been corrected by subtracting a linear term $\chi_d H$, where $\chi_d$ is the high-field slope of the raw data. The $\chi_d$ values are practically identical for all **H** orientations, as expected for the cubic STO substrate (not shown). Note that the presence of such background, however, would conceal any possible contribution from non-saturating behavior of the CFO film at high fields, as commonly observed in these systems [54,59,60].

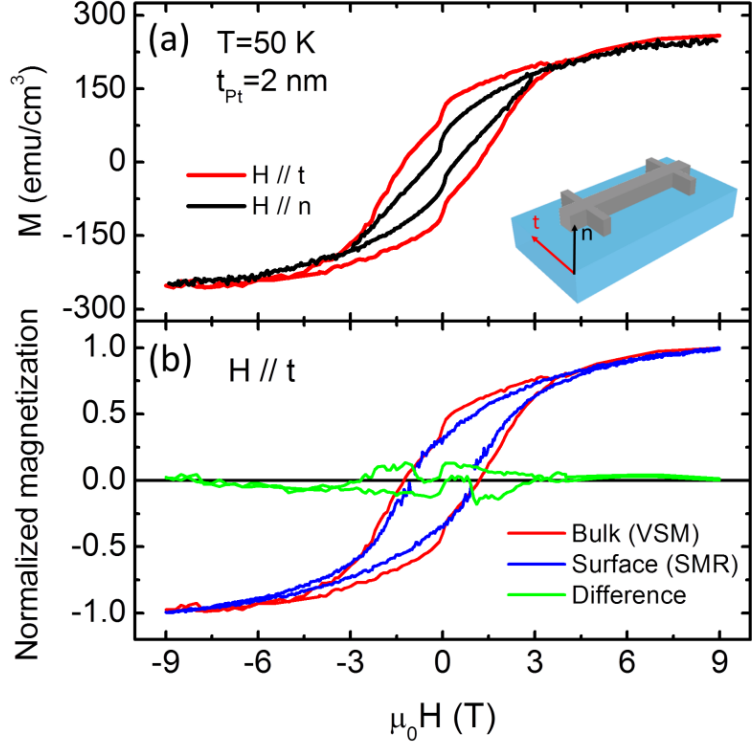

**Figure 5**. *(a) Magnetic hysteresis loops for the Pt(2 nm)/CFO(40 nm) sample measured by VSM at 50 K. H is applied along **t** (red curve) and **n** (black curve), as defined in the inset. A diamagnetic background has been subtracted (b) Comparison of normalized hysteresis loops with H applied along **t** for the same sample. Red curve is the one in (a), normalized to the maximum value. Blue curve is computed from the longitudinal resistance $R_L$ at 50 K as a function of H applied along **t**, which is shown in Fig. 3c. Green curve is the difference between the two hysteresis loops.*

Figure 3c shows the longitudinal resistance of the Pt(2 nm)/CFO(40 nm) sample measured at different orthogonal **H** orientations ($R_L(\mathbf{H}\|\mathbf{j})$, $R_L(\mathbf{H}\|\mathbf{t})$ and $R_L(\mathbf{H}\|\mathbf{n})$), after subtracting the background resistance. When applying a field $\mathbf{H}\|\mathbf{n}$, the transverse component of the magnetization $m_t$ should be reduced and, following Eq. 1, $R_L(\mathbf{H}\|\mathbf{n})$ should increase with H. A similar behavior is expected for $R_L(\mathbf{H}\|\mathbf{j})$ when applying a field $\mathbf{H}\|\mathbf{j}$. Accordingly, $R_L(\mathbf{H}\|\mathbf{t})$ should decrease with increasing H. The experimental data in Fig. 3c confirms these trends. As expected, this behavior shares similarities and is reminiscent of the M(H) curves in Fig. 5a because, as shown by Eq. 1, the field evolution of M(H) should translate into $R_L(H)$. Taking into account that $m_t$ in Eq. 1 is the cosine director of the magnetization **M** along the **t** direction ($m_t = \cos(\Psi)$), one could perform a more quantitative analysis by extracting the angle $\Psi$ between **M** and **t** as a function of the applied magnetic field from $R_L(\mathbf{H}\|\mathbf{t})$. In particular, Eq. 1 can be rewritten as $(\rho_L-\rho_0)/\Delta\rho_1 = (1-m_t^2) = \sin^2(\Psi)$. In order to normalize our magnetoresistance measurements in the same way, we assume that the local maxima $R_{max}$ in the $R_L(\mathbf{H}\|\mathbf{t})$ curve (see Fig. 3c) corresponds to $\Psi=90°$ and the value at the largest field (9 T) is saturated and corresponds to $\Psi=0°$. These assumptions are not strictly correct because: i) we do not know exactly how **M** rotates with **H** (we could have a $\Psi$ lower than 90° in the local maxima) and ii) $R_L(\mathbf{H}\|\mathbf{t})$ does not fully saturate at 9 T. With these precautions, we compute $[R_L(H)-R_L(9\ T)]/[R_{max}-R_L(9\ T)] \approx \sin^2(\Psi)$. From here, one can calculate $m_t=\cos(\Psi)$, which is also the **M** component measured in the VSM magnetometer when using the in-plane field configuration, as a function of the magnetic field. Note that the above analysis cannot be done for $R_L(\mathbf{H}\|\mathbf{j})$ and $R_L(\mathbf{H}\|\mathbf{n})$ because in those cases the other two components of **M** ($m_n$ and $m_j$) cannot be univocally determined.

Figure 5b shows the resulting hysteresis loop (blue line). Superposed is the normalized M(H) data obtained with VSM magnetometry along the **t** direction (red line). It is remarkable that the shape of the hysteresis loop extracted from the surface-sensitive SMR measurements is similar to the bulk-sensitive M(H) loop, except for the characteristic steps observed around zero field in the M(H) curve which are absent in $m_t$(H). Accordingly, our semi-quantitative analysis strongly suggests a reduced density of APBs at the surface of our CFO films while APBs are present in the bulk. There is abundant literature indicating that the density of APBs in spinel oxide films ($Fe_3O_4$ [57] and CFO [63]) decreases as the films get thicker. On the basis of these results it could be expected that the density of APBs also decreases when approaching the film surface. Indeed, APBs are defective regions generated essentially at the first layers during initial growth and their presence unavoidably produces strain gradients in the films. Elastic energy shall be released during film growth and the simplest way is to reduce the APBs density away from the interface with the substrate. Within the scope of APBs model developed by Sofin *et al.* [62], the observed difference between bulk and surface magnetization (green line in Fig. 5b) would reflect the magnetization reversal of domains antiferromagnetically coupled across APBs.

From this comparison, we conclude that SMR is extremely sensitive to fine details of the magnetic ordering at the Pt/CFO interface and $R_L(H)$ data are thus fingerprints of a distinct surface magnetization which, although not discernible in the bulk-sensitive magnetization experiments, largely dominates the longitudinal SMR. A further systematic SMR study in CFO films with different thicknesses will help understanding the particular complexity of surface magnetization in spinel structures.

**IIId. Surface magnetization as a function of CFO thickness**

For this reason, we fabricated *ex-professo* a new set of Pt/CFO samples with a fixed Pt thickness of 2 nm (in order to maximize the SMR amplitude) and CFO thicknesses of 20, 30, 40 and 60 nm. We systematically measured hysteresis loops with VSM for all thicknesses at temperatures between 10 and 300 K [64]. As mentioned above, the approach to saturation critically depends on the protocol to subtract the diamagnetic background [64]. Most of the corrected hysteresis loops display a saturated behavior as a linear contribution has been subtracted, but also some inconsistencies are present (see Fig. S1 and 7), i.e., information on a high field magnetic susceptibility is lost by this data treatment, which thus invalidates any critical analysis of the approach to saturation. We also stress that, in this set of samples, the characteristic steps at zero field in the magnetization curves are only observed clearly for the thinnest (20 nm) CFO film (Fig. S1 and 7); this finding is fully consistent with the proposed inverse correlation between APB density and film thickness.

FDMR has been measured at several temperatures (50, 150, 300 K) with the magnetic field applied in two relevant directions (**H**||**n** and **H**||**t**). Figure 6 shows the results for two different CFO thicknesses (20, 60 nm). The amplitudes of SMR are very similar to those observed in the Pt(2 nm)/CFO (40 nm) sample of the previous set (Fig. 2c).

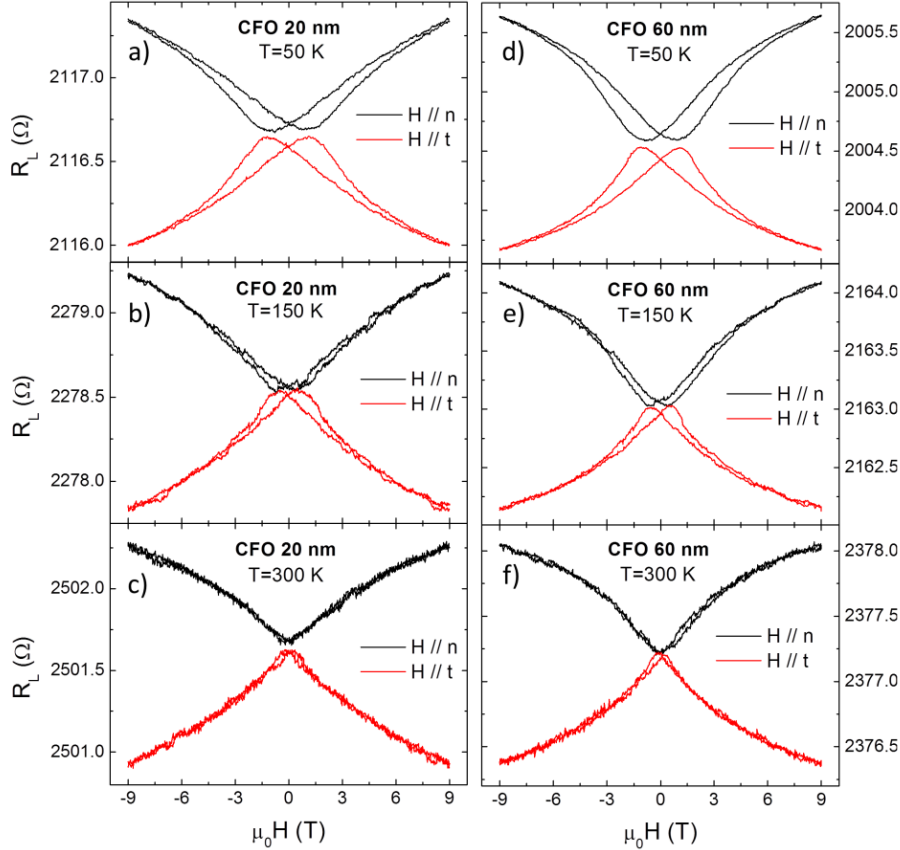

FIG 6. *Longitudinal resistance $R_L$ for (a-c) Pt(2 nm)/CFO(20 nm) and (d-f) Pt(2 nm)/CFO(60 nm) samples, as a function of H applied along **t** (red curves) and **n** (black curves). Measurements were done at 50 K (a,d), 150 K (b,e) and 300 K (c,f).*

We use the semi-quantitative model described in the previous subsection to derive a normalized hysteresis loop for Pt(2 nm)/CFO(20 nm) and Pt(2 nm)/CFO(60 nm) samples at different temperatures. In Fig. 7, we show the comparison of the SMR hysteresis loops with the normalized hysteresis loops obtained by VSM. Two main features should be highlighted here: i) In all cases, SMR hysteresis loops display a slow approach to saturation. We stress again that this information could not be conclusively assessed from the VSM hysteresis loops. ii) The characteristic steps at zero field in the VSM loop for Pt(2 nm)/CFO(20 nm) are not present in the corresponding SMR loop (Fig. 7a), confirming our previous observation from Fig. 5b.

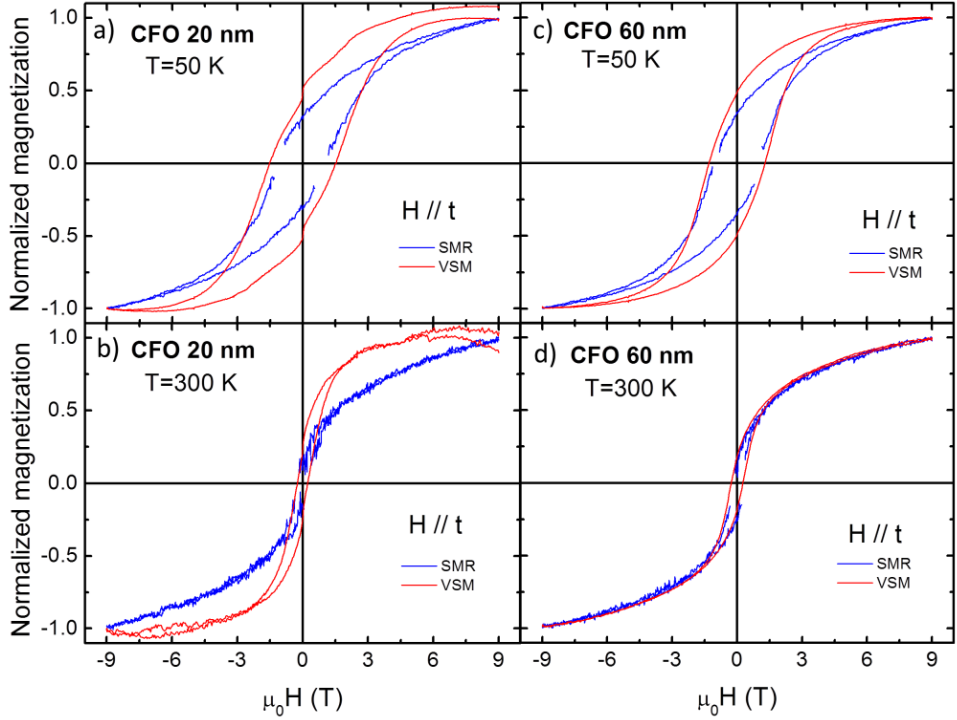

FIG. 7. *Comparison of normalized magnetic hysteresis loops with H applied along **t** for (a,b) Pt(2 nm)/CFO(20 nm) and (c,d) Pt(2 nm)/CFO(60 nm) samples at 50 K (a,c) and 300 K (b,d). Red curve is measured by VSM magnetometry and normalized to the maximum value after diamagnetic background subtraction. Blue curve is computed from the longitudinal resistance $R_L$ as a function of H applied along **t**, which is shown in Fig. 6.*

In Fig. 8, we plot SMR hysteresis loops for Pt(2 nm)/CFO(20 nm) and Pt(2 nm)/CFO(60 nm) samples at different temperatures. Surprisingly, the SMR hysteresis loops are almost identical for 20- and 60-nm-thick CFO films at any temperature. This is not the case for the VSM hysteresis loops. This impressive result clearly indicates that the surface has the same magnetic behavior independently of the CFO film thickness, which is completely different from the bulk behavior, and highlights the importance of surface anisotropy.

In summary, this series of hysteresis loops derived from SMR for different temperatures and CFO thicknesses gives us consistent results on the complex magnetism of CFO, impossible to obtain with any standard magnetometry technique: i) the surface magnetization behaves independently of the CFO thickness and does not saturate up to 9 T, evidencing that the surface has its own anisotropy; ii) the characteristic zero-field steps are not present at the surface magnetization while they are relevant in the bulk, strongly suggesting that APBs are the responsible of such intriguing features.

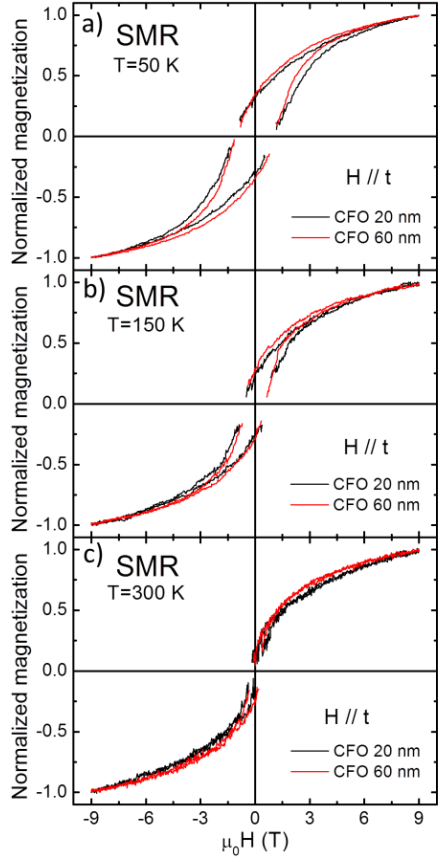

FIG. 8. *Comparison of normalized hysteresis loops for Pt(2 nm)/CFO(20 nm) (black curves) and Pt(2 nm)/CFO(60 nm) (red curves) samples, computed from the longitudinal resistance $R_L$ at (a) 50 K, (b) 150 K and (c) 300 K, as a function of H applied along **t** (shown in Fig. 6).*

## IV. Conclusions

We have reported longitudinal magnetoresistance measurements of *in-situ* grown Pt on ferromagnetic insulating CFO, as a function of intensity and orientation of the magnetic field. We have shown that even if the data can be well described by the spin Hall magnetoresistance, an additional feature appears at high magnetic fields. This additional effect originates from the ordinary magnetoresistance of Pt. Furthermore, we have compared the field dependent longitudinal resistance arising from the SMR to the bulk magnetization of the CFO. This analysis reveals important differences that we correlate to the significant role of antiphase boundaries and surface anisotropy in systems such as spinels where strong competition of magnetic interactions takes place. This shows the tremendous potential of spin Hall magnetoresistance to probe the surface magnetization of ferromagnetic insulators, otherwise not possible with standard magnetometric techniques.

## Acknowledgments

This work is supported by the European Union under the NMP project (263104-HINTS), and the European Research Council (257654-SPINTROS), by the Spanish MINECO (MAT2012-37638, MAT2015-65159-R and MAT2014-56063-C02), by the Basque Government (PC2015-1-01) and by Generalitat de Catalunya (2014 SGR 00734). M. I. acknowledges the Basque Government for a PhD fellowship (BFI-2011-106). J. F. acknowledges stimulating discussions with Xavier Martí.

**Spin Hall magnetoresistance as a probe for surface magnetization in Pt/CoFe$_2$O$_4$ bilayers**

Miren Isasa[1], Saül Vélez[1], Edurne Sagasta[1], Amilcar Bedoya-Pinto[1], Nico Dix[2], Florencio Sánchez[2], Luis E. Hueso[1,3], Josep Fontcuberta[2] and Fèlix Casanova[1,3]

[1]CIC nanoGUNE, 20018 Donostia-San Sebastian, Basque Country, Spain.
[2]Institut de Ciència de Materials de Barcelona (ICMAB-CSIC), Campus UAB, 08193 Bellaterra, Catalonia, Spain.
[3]IKERBASQUE, Basque Foundation for Science, 48013 Bilbao, Basque Country, Spain.

# SUPPLEMENTAL MATERIAL

In any magnetic thin film having a large magnetic anisotropy (such as CoFe$_2$O$_4$ containing Co$^{2+}$ ions), surface anisotropy effects (and the associated magnetic hardness) and the unavoidable contribution of the substrate to the measured magnetization makes the study of the approach to saturation rather challenging. It is in this region where most commonly surface effects dominate. Bulk magnetometry cannot disentangle substrate and film contributions and thus we are blind to surface magnetization properties among other aspects.

This is dramatically clear in Fig. S1: Figs. S1a and S1b show the raw data of the hysteresis loops obtained by VSM magnetometry for the second set of samples [Pt (2 nm)/CFO(20, 30, 40 and 60 nm)] at two different temperatures (50 K and 300 K). Important parameters such as the saturation magnetization and coercive field critically depend on the subtraction protocol. The protocol we have chosen is to subtract, for each sample, the diamagnetic background at 10 K and use it for all temperatures (the diamagnetic contribution should be temperature independent), obtaining Figs. S1c and S1d. In this case, both the saturation magnetization (Fig. S1e) and the coercive field (Fig. S1f) vary systematically with CFO thickness and temperature. Most of the corrected hysteresis loops display a saturating behavior as a linear contribution has been subtracted (see for instance Fig. S1c), but also some inconsistencies are present (see high field regions in Fig. S1d): i.e., information on a high field magnetic susceptibility is lost by this data treatment, which thus invalidates any critical analysis of the approach to saturation. We also stress that the characteristic steps at zero field in the magnetization curves are only observed clearly for the thinnest (20 nm) CFO film; this finding is fully consistent with the proposed inverse correlation between antiphase boundaries (APB) density and film thickness.

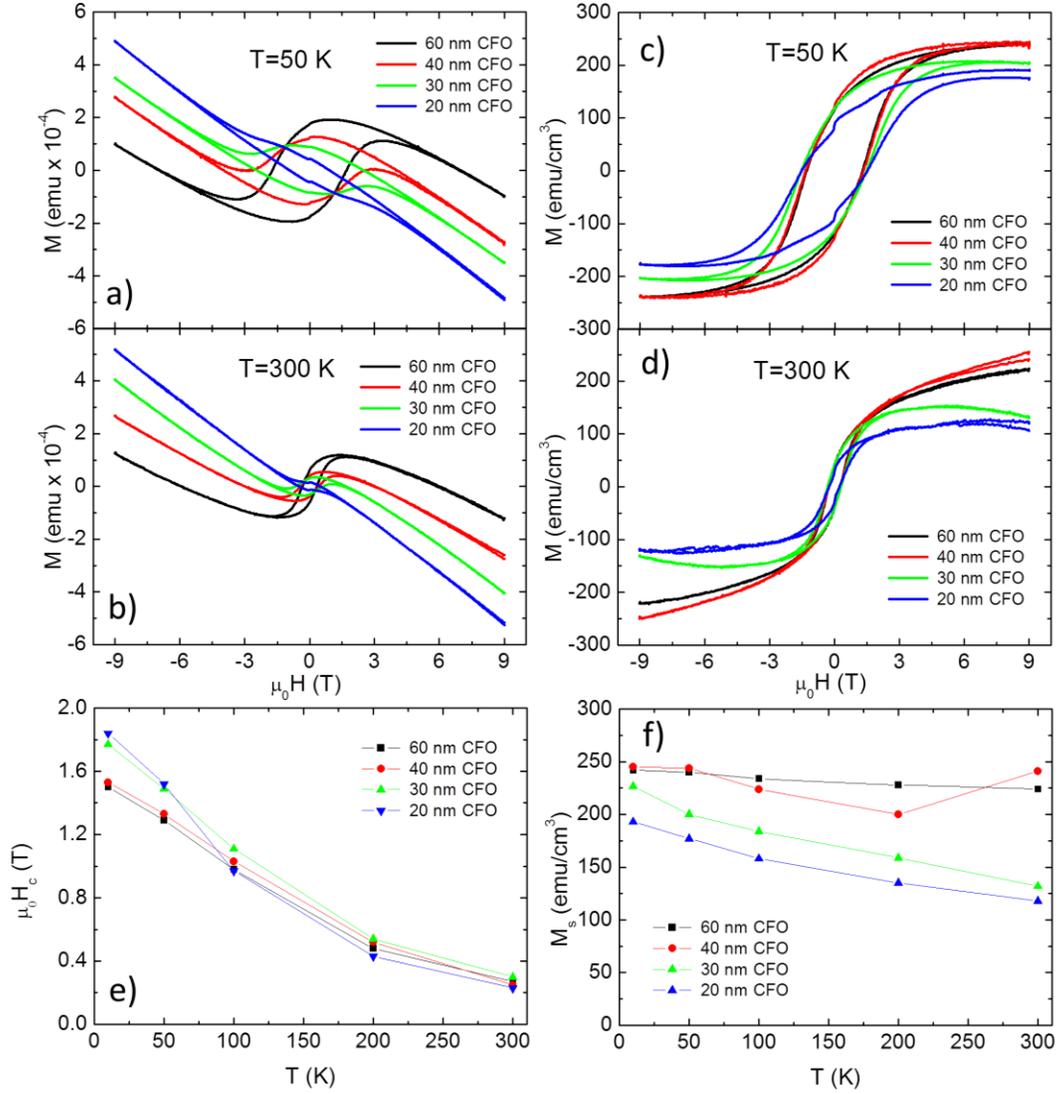

FIG. S1. *Raw data of magnetic hysteresis loops for the Pt(2 nm)/CFO(20, 30, 40, 60 nm) samples measured by VSM at a) 50 K and b) 300 K. Magnetic field H is applied along **t**, as defined in the main text. Hysteresis loops at c) 50 K and c) 300 K for the same samples after a diamagnetic background has been subtracted and the corrected data have been normalized by the CFO volume. e) Coercive field and f) saturation magnetization as a function of temperature for the same samples, extracted from the corrected hysteresis loops.*